\documentclass[journal]{IEEEtran}

\usepackage{amsmath}
\usepackage{amssymb}
\usepackage{bm}
\usepackage[utf8]{inputenc}

\begin{document}

\title{Exponential Family Models from Bayes' Theorem under Expectation Constraints}

\author{Sergio Davis
\thanks{The author is with the Comisión Chilena de Energía Nuclear, Casilla
188-D, Santiago, Chile.}}

\markboth{IEEE Transactions on Information Theory,~Vol.~XXX, No.~XXX, XXX}%
{Davis: Exponential family models from Bayes' theorem under expectation
constraints}

\maketitle

\begin{abstract}
It is shown that a consistent application of Bayesian updating from a prior probability 
density to a posterior using evidence in the form of expectation constraints leads to exactly the 
same results as the application of the maximum entropy principle, namely a posterior belonging to the 
exponential family. The Bayesian updating procedure presented in this work is not expressed as a variational 
principle, and does not involve the concept of entropy. Therefore it conceptually constitutes a complete 
alternative to entropic methods of inference.
\end{abstract}

\section{Introduction}

The principle of maximum entropy (MaxEnt for short) is considered a fundamental tool in the 
application of information theory to the construction of probabilistic models~\cite{Presse2013}. 
First proposed in the context of Statistical Mechanics by J. W. Gibbs~\cite{Gibbs1902} as a derivation 
of the canonical ensemble from the requirement of fixed average internal energy, it was recast by 
Jaynes~\cite{Jaynes1957} as a \emph{general principle of reasoning} following
the insights of Shannon~\cite{Shannon1948}.

This principle of reasoning under uncertainty, that we call MaxEnt, postulates that the most unbiased 
probability assignment $\rho^*(\bm u)$ among those probability densities $\rho(\bm u)$ consistent 
with given information $E$, is the one that maximizes the universal \textbf{entropy} functional

\begin{equation}
\mathcal{S}[\rho; \rho_0] = -\int_U d^n u\; \rho(\bm u)\ln \frac{\rho(\bm u)}{\rho_0(\bm u)},
\label{eq_shannon}
\end{equation}
under the constraints imposed by $E$. Here $U$ is an $n$-dimensional state space and $\bm{u} \in U$ 
a possible state of the system. The density $\rho_0(\bm{u})$ is the initial probability density 
representing complete ignorance about the value of $\bm{u}$.

The maximization problem is solved using the method of Lagrange multipliers, and for the case of information 
$E$ expressed as the given expectation $\Big<f(\bm{u})\Big>=F$ the solution is well-known~\cite{Jaynes1957,Jaynes2003}, 
and corresponds to the standard \emph{exponential family} of models,

\begin{equation}
\rho^*(\bm{u}) = \frac{1}{Z(\lambda)}\rho_0(\bm{u})\exp(-\lambda f(\bm{u}))
\label{eq_canonical}
\end{equation}
where $\lambda$ is the Lagrange multiplier which can be determined from the implicit equation

\begin{equation}
-\frac{\partial}{\partial \lambda}\ln Z(\lambda)=F.
\label{eq_lambda}
\end{equation}
 
The uniqueness of the MaxEnt procedure relies on the uniqueness of the entropy functional $\mathcal{S}$ 
of Eq. \ref{eq_shannon}, and this has been established from consistency requirements~\cite{Shore1980, Skilling1988, 
Garrett2001, Caticha2006}. These axiomatic approaches have allowed a useful separation between the operational 
formalism of MaxEnt updating (in the sense of formulating an optimal inference rule) and the meaning of the 
entropy $\mathcal{S}$ being maximized. 

In the purely operational vision, entropy is just a convenient device which allows the ranking 
of the different candidate distributions from best to worst, and we do not need to assign any meaning to it. 
However, this ranking must, at some point, be understood in terms of some quality of the candidate models in 
terms of which they are rejected or accepted. According to the standard interpretation of the entropy $\mathcal{S}$ by 
Shannon (later adopted by Jaynes), the quantity $\mathcal{S}$ is a measure of the \emph{missing information} 
needed to completely determine the state $\bm{u}$. In maximizing $\mathcal{S}$ we are choosing the least informative 
model that agrees with all our constraints.

The original entropy axioms by Shannon~\cite{Shannon1948} as well as the Shore and Johnson axioms~\cite{Shore1980} 
have been recenly put into question~\cite{Tsallis2015} because they rule out the so called generalized entropies 
(such as the Tsallis or Renyi entropies) for use in inference~\cite{Presse2013b}. In this ongoing development it 
would be highly desirable to have alternative methods for updating probabilities, and an obvious candidate is Bayes' 
theorem itself. Both MaxEnt and Bayes' theorem solve essentially the same problem, namely updating a prior to a 
posterior incorporating some new information or evidence, regardless of the nature of that evidence (measured data 
or model constraints). Looking closely, we can in fact recognize the exponential form given in Eq.~\ref{eq_canonical} 
as a Bayesian updating from a \emph{prior} distribution $\rho_0(\bm{u})$ to a posterior distribution 
$\rho^*(\bm{u})=M(\bm{u}) \rho_0(\bm{u})$  with an ``updating factor'' 

\begin{equation}
M(\bm{u}) \propto \exp\big(-\sum_i \lambda_i f_i(\bm{u})\big).
\label{eq_M_factor}
\end{equation}
 
If we can arrive at this form of $M$ without invoking a maximization of some functional, then we can effectively 
bypass the issue of adequacy of $\mathcal{S}$. In a frequentist context (taking expectations as the limit of statistical 
averages) Campenhout and Cover~\cite{Campenhout1981} have produced this factor $M$ in the limit of infinite samples.
 
In this work we provide a general derivation of the update rule in Eq. \ref{eq_canonical} from Bayesian inference 
subjected to expectation constraints. Unlike the result by Campenhout and Cover, the proof does not depend on frequentist 
assumptions such as the identification of expectations with averages over samples, or even the assumption of data samples being processed.
It is based simply on imposing consistency conditions (in similar spirit to Cox~\cite{Cox1961} and Shore and Johnson) on the Bayesian updating 
rules themselves to constrain the form of $M$.

The work is organized as follows. First, in section \ref{sec_notation}, we present the notation to be used in 
the rest of the paper and give an outline of the assumptions needed for the central claim of the paper.
In sections \ref{sec_repre}, \ref{sec_cvt} and \ref{sec_indep}, the exponential form is obtained from the assumptions. 

\section{Mathematical notation and statement of the derivation}
\label{sec_notation}

We will consider in the following a system described by states $\bm{u}$ in an
$n$-dimensional state space $U$. We start from an arbitrary state of knowledge
$I_0$ with probability density $P(\bm{u}|I_0)$ and our goal is to perform a
Bayesian update to a posterior density $P(\bm{u}|I)$ where $I=I_0\wedge E$ is the new
state of knowledge that includes the evidence $E$. We will denote the expectation of a 
quantity $A(\bm{u})$ under the state of knowledge $J$ by $\big<A\big>_J$,
defined as\footnotemark

\begin{equation}
\Big<A\Big>_J = \int_U d^nu\; P(\bm{u}|J) A(\bm{u}).
\end{equation}

\footnotetext{
Note that, if $A$ is a scalar (i.e. invariant under a change of coordinate
system) then $\big<A\big>_J$ is also a scalar, because $P(\bm{u}|J)$ is a scalar
density, which transforms just as the invariant measure $\sqrt{g(\bm{u})}$.
Therefore we do not include $\sqrt{g}$ explicitly in the integral.}

For the problem of inference with prior $P(\bm{u}|I_0)$ and evidence $E$ given
by $\big<f(\bm{u})\big>_I=F$, we define the following \textbf{consistency conditions}:

\vspace{10pt}
\begin{enumerate}
\item [(a)] The probability ratio $P(E|\bm{u},I_0)/P(E|I_0)$ is a unique functional of the constraining
function $f$, evaluated using the state $\bm{u}$ and the constraining value $F$. That is, we can write 

\begin{equation}
\frac{P(E|\bm{u}, I_0)}{P(E|I_0)} = M[f](\bm{u}, F).
\end{equation}

The functional $M$ encodes the method of inference, and we are looking for a unique method. The only 
information we have in using this method is the evidence $E$, which consists of the function $f$ and 
its expectation $F$, so $M$ cannot depend on any other piece of information, such as additional parameters.

\vspace{10pt}
\item [(b)] The inference considering $f$ as a function of $\bm{u}$ is
consistent with the inference considering $f$ as a fundamental variable. In
other words, it should be possible to ignore the degrees of freedom $\bm{u}$ and perform an
inference over $f$ itself, using the same functional $M$. This inference has to
be valid and consistent with the full inference using $\bm{u}$. 

\vspace{10pt}
\item [(c)] Logically independent subsystems $U_1$ and $U_2$ can be analyzed separately or 
jointly as $U=U_1\otimes U_2$, producing the same result. This \textbf{does not} restrict the method 
to separable systems only, it only ensures that \textbf{if} we decide to apply it to a pair of independent 
systems, the method should preserve their independence. The same functional $M$ must be valid for systems 
of arbitrary correlation.

\end{enumerate}

In the following we will show that these conditions imply a posterior 

\begin{equation}
P(\bm{u}|E,I_0) = \frac{1}{Z(\lambda)}\exp(-\lambda f(\bm{u}))P(\bm{u}|I_0).
\label{eq_expform}
\end{equation}
 
Here $\lambda$ is a parameter to be fixed by the value of $F$, not a Lagrange multiplier, as there 
is no variational procedure in the derivation.

The proof is divided in two parts. First, we prove that conditions (a) and (b)
imply that the functional $M[f](\bm{u}, F)$ is actually a function of two arguments, $m(f(\bm{u}), F)$.
Finally, we show that condition (c) implies $m(f, F) \propto \exp(-\lambda(F)f)$, which immediately 
leads to Eq. \ref{eq_expform} after normalization.

\section{Consistency between different representations}
\label{sec_repre}

The evidence $E$, consisting of the given expectation $\big<f\big>_I=F$, can be
used in two different ways. If we regard the quantity $f$ as a variable in
itself, it can be used to update the prior $P(f|I_0)$ to a posterior $P(f|I)$, given by 

\begin{equation}
P(f|I) = P(f|I_0)M[\mathbb{I}](f, F).
\label{eq_prob_f_1}
\end{equation}

Here the functional depends on the identity function $f \rightarrow f$, denoted by $\mathbb{I}$. 
On the other hand, we can take $f$ as a function of the degrees of freedom $\bm{u}$, and use the
evidence $E$ to update the prior $P(\bm{u}|I_0)$ to a posterior $P(\bm{u}|I)$,

\begin{equation}
P(\bm{u}|I) = P(\bm{u}|I_0)M[f](\bm{u}, F).
\label{eq_prob_f_2}
\end{equation}
 
For both problems the same functional $M$ should be used and yield consistent
posteriors. Through the laws of probability, the probability density of $f$ is 
always connected to the probability density of $\bm{u}$ by

\begin{equation}
P(f|J) = \Big<\delta(f(\bm{u})-f)\Big>_J,
\end{equation}
for every state of knowledge $J$, which in our case produces two independent relations, 
for $J=I$ and $J=I_0$, namely

\begin{eqnarray}
P(f|I) = \int d^nu\; P(\bm{u}|I)\delta(f(\bm{u})-f), \\
P(f|I_0) = \int d^nu\; P(\bm{u}|I_0)\delta(f(\bm{u})-f).
\label{eq_bar}
\end{eqnarray}

By replacing Eqs. \ref{eq_prob_f_1} and \ref{eq_prob_f_2} we find a constraint for $M$,
namely

\begin{equation}
\int d^nu\; P(\bm{u}|I_0)\delta(f(\bm{u})-f)\Big[M[f](\bm{u}, F)-M[\mathbb{I}](f, F)\Big] = 0
\end{equation}
for any prior $P(\bm{u}|I_0)$. The delta function allows us to replace $f$ by
$f(\bm{u})$ in the arguments to $M[\mathbb{I}]$, so we have

\begin{equation}
\int d^nu\; P(\bm{u}|I_0)\delta(f(\bm{u})-f)\Big[M[f](\bm{u}, F)-M[\mathbb{I}](f(\bm{u}), F)\Big] = 0.
\end{equation}

Then, taking the functional derivative on both sides with respect to the prior, we have 

\begin{equation}
M[f](\bm{u}, F) = M[\mathbb{I}](f(\bm{u}), F) = m(f(\bm{u}), F).
\end{equation}
 
This means that for any constraining function $f$ the functional $M$ reduces to a function of 
$f(\bm{u})$ (the constraining function evaluated at the state $\bm{u}$) and $F$ (the constraining value); 
it does not, for instance, depend on derivatives of $f$. We have then constrained the form of the posterior 
distribution to be

\begin{equation}
P(\bm{u}|I) = P(\bm{u}|I_0)m(f(\bm{u}), F)
\label{eq_posterior_m}
\end{equation}
with $m$ a unique function, to be determined. 

\section{A theorem for posterior expectations}
\label{sec_cvt}

In order to find this universal function $m(f, F)$ we first present a useful
identity between expectations. Applying Stoke's theorem to an arbitrary probability 
density $P(\bm{u}|J)$ we obtain (see the Appendix) the following expectation identity,

\begin{equation}
\Big<\nabla \cdot \bm{v}\Big>_J + \Big<\bm{v}\cdot\nabla \ln P(\bm{u}|J)\Big>_J = 0,
\end{equation}
valid for any differentiable vector function $\bm{v}(\bm{u})$. This is a generalization of the result in Ref.~\cite{Davis2012}. 

Replacing the form of the posterior found in Eq. \ref{eq_posterior_m}, we have,
for the state of knowledge $I$,

\begin{equation}
\Big<\nabla \cdot \bm{v}\Big>_I + \Big<\bm{v}\cdot\nabla \ln
P(\bm{u}|I_0)\Big>_I = -\Big<\bm{v}\cdot \nabla \ln m(f(\bm{u}), F)\Big>_I,
\end{equation}
or, more compactly,

\begin{equation}
\Big<\mathcal{D}_0 \bm{v}\Big>_I = \Big<\hat{\lambda}(f(\bm{u}), F)\bm{v}\cdot\nabla f\Big>_I,
\end{equation}
where the operator $\mathcal{D}_0$ is defined as $\mathcal{D}_0 \bm{v}=\nabla \cdot \bm{v}+\bm{v}\cdot\nabla \ln P(\bm{u}|I_0)$ 
and 

\begin{equation}
\hat{\lambda}(f, F)= -\frac{\partial}{\partial f}\ln m(f, F).
\label{eq_hat_lambda}
\end{equation}

In the next section we will constrain the form of this function $\hat{\lambda}$
using the condition (c) on independent systems. We will show that $\hat{\lambda}$ only 
depends on its second argument $F$.
 
\section{Consistency between separate and joint treatment of independent subsystems}
\label{sec_indep}

Now let us consider the following situation: two logically independent systems, $U_1$ and $U_2$ 
with priors $P(\bm{u}_1|I_0)$ and $P(\bm{u}_2|I_0)$ respectively. We can decide
to update these priors separately using the evidence $E$ given by

\begin{eqnarray}
\Big<f_1(\bm{u}_1)\Big>_I = F_1 \label{eq_constraint1}, \\
\Big<f_2(\bm{u}_2)\Big>_I = F_2.
\label{eq_constraint2}
\end{eqnarray}
 
We could also decide to update the joint system prior $P(\bm{u}|I_0)=P(\bm{u}_1|I_0)P(\bm{u}_2|I_0)$,
using the evidence

\begin{equation}
\Big<f_1(\bm{u}_1)+f_2(\bm{u}_2)\Big>_I = F_1+F_2.
\label{eq_constraint_combined}
\end{equation}

In both cases we are forced, by condition (a), to use Eq. \ref{eq_posterior_m}
with the same function $m$ on each case. From Eqs. \ref{eq_constraint1} and \ref{eq_constraint2} 
applied separately on $\bm{u}_1$ and $\bm{u}_2$, we obtain 

\begin{eqnarray}
\Big<\mathcal{D}_0 \bm{v}_1(\bm{u}_1)\Big>_I = \Big<\hat{\lambda}(f_1(\bm{u}_1), F_1)\bm{v}_1\cdot\nabla f_1\Big>_I, \\
\Big<\mathcal{D}_0 \bm{v}_2(\bm{u}_2)\Big>_I = \Big<\hat{\lambda}(f_2(\bm{u}_2), F_2)\bm{v}_2\cdot\nabla f_2\Big>_I,
\end{eqnarray}

and from Eq. \ref{eq_constraint_combined},

\begin{equation}
\Big<\mathcal{D}_0 \bm{v}(\bm{u})\Big>_I = \Big<\hat{\lambda}(f_1(\bm{u}_1)+f_2(\bm{u}_2), F_1+F_2)\bm{v}\cdot\nabla f\Big>_I.
\end{equation}

Choosing first $\bm{v}(\bm{u}_1)=\bm{v}_1$ and then $\bm{v}(\bm{u}_2)=\bm{v}_2$ we find (calling $f_{12}=f_1+f_2$),

\begin{eqnarray}
\Big<\hat{\lambda}(f_{12}, F_1+F_2)\bm{v}_1\cdot\nabla f_1\Big>_I
= \Big<\hat{\lambda}(f_1, F_1)\bm{v}_1\cdot\nabla f_1\Big>_I, \\
\Big<\hat{\lambda}(f_{12}, F_1+F_2)\bm{v}_2\cdot\nabla f_2\Big>_I
= \Big<\hat{\lambda}(f_2, F_2)\bm{v}_2\cdot\nabla f_2\Big>_I,
\end{eqnarray}
for every choice of $\bm{v}_1$ and $\bm{v}_2$. This implies 

\begin{equation}
\hat{\lambda}(f_{12}, F_1+F_2) = \hat{\lambda}(f_1, F_1) = \hat{\lambda}(f_2, F_2).
\end{equation}

As $f_1$ only depends on $\bm{u}_1$, and $f_2$ only depends on $\bm{u}_2$, the last
equality means that, in general, the function $\hat{\lambda}(f, F)$ does not
depend on $f$; it is only a function of the second argument $F$, so that we can
replace $\hat{\lambda(f(\bm{u}), F)}$ by $\lambda(F)$ such that

\begin{equation}
\lambda(F_1+F_2) = \lambda(F_1) = \lambda(F_2).
\end{equation}

Replacing $\hat{\lambda}(f(\bm{u}), F)$ by $\lambda(F)$ into the definition of $\hat{\lambda}$ 
(Eq. \ref{eq_hat_lambda}), we have 

\begin{equation}
-\frac{\partial}{\partial f}\ln m(f, F) = \lambda(F),
\end{equation}
and therefore,

\begin{equation}
m(f, F) = m_0(F)\exp(-\lambda(F)f).
\end{equation}

The posterior distribution for $\bm{u}$ in Eq. \ref{eq_posterior_m} then reads,

\begin{equation}
P(\bm{u}|I) = m_0(F)\exp(-\lambda(F)f(\bm{u}))P(\bm{u}|I_0),
\end{equation}
with $m_0(F)$ fixed by normalization to be

\begin{equation}
\frac{1}{m_0(F)} = \int d^nu\; \exp(-\lambda(F)f(\bm{u}))P(\bm{u}|I_0) = Z(\lambda(F)).
\end{equation}

As all the dependence on $F$ is through $\lambda(F)$, we can finally write the 
distribution entirely as a function of $\lambda$, so 

\begin{equation}
P(\bm{u}|I) = \frac{1}{Z(\lambda)}\exp(-\lambda f(\bm{u}))P(\bm{u}|I_0),
\end{equation}
with $\lambda=\lambda(F)$ a number to be determined to agree with the constraint
$\big<f(\bm{u})\big>_I = F$. Given this functional form of $P(\bm{u}|I)$ we can 
write the expectation of $f$ as a derivative of $\ln Z(\lambda)$, 

\begin{equation}
\Big<f(\bm{u})\Big>_I = -\frac{\partial}{\partial \lambda}\ln Z(\lambda)
\end{equation}
and then it follows that the constraint fixes $\lambda$ through Eq. \ref{eq_lambda}, as expected.

\section{Conclusion}

We have proved that Bayesian updating given an expectation constraint can be established as a uniquely 
defined procedure, leading to an exponential family posterior density, which is the same result produced 
by the application of the principle of maximum entropy (MaxEnt) under the same constraints. 
We only require that a unique ``updating factor'' is used in all cases, and that its use is consistent 
through different definitions of the state space. That the same answer is revealed using an alternative method 
of inference shows that the essence of MaxEnt is already contained in Bayes' theorem under our additional 
consistency requirements. This allows a conceptual unification of the fields of Bayesian inference and MaxEnt 
inference under a common framework for continuous degrees of freedom. Our derivation also shows that the 
existence of an entropy functional is not central to the core of inference; although it certainly provides 
a more than convenient device for practical calculations, it is conceptually not required for a consistent 
formulation of a theory of inference. 

\appendix[An identity for expectations of continuous variables]

We will consider an $n$-dimensional manifold $U$ with metric tensor $g_{\mu\nu}(\bm{u})$ and induced 
metric $h_{\mu\nu}(\bm{u})$ on the surface $\partial U$ which acts as a boundary of $U$. Recall the 
covariant form of Stoke's theorem~\cite{Carroll2004},

\begin{equation}
\int_U d^nu \sqrt{g}\; \nabla_\mu \omega^\mu = \int_{\partial U}d^{n-1}u \sqrt{h}\; n_\sigma \omega^\sigma,
\label{eq_divergence}
\end{equation}
with $\bm{\omega}$ any differentiable field, $\bm{n}$ the normal to $\partial U$ and 
$\nabla_\mu \omega^\mu$ the covariant divergence, defined as

\begin{equation}
\nabla_\mu \omega^\mu = \partial_\mu \omega^\mu + \omega^\mu\partial_\mu \ln \sqrt{g}.
\end{equation}

Now consider the field 

\begin{equation}
\omega^\mu=\frac{P(\bm{u}|J)}{\sqrt{g}}v^\mu(\bm{u}),
\end{equation}
where the probability density $P(\bm{u}|J)$ vanishes in the boundary $\partial
U$. Replacing in Eq. \ref{eq_divergence}, we have

\begin{eqnarray}
\int_U d^nu P(\bm{u}|J)\Big(\partial_\mu v^\mu + v^\mu\partial_\mu \ln P(\bm{u}|J)\Big) \nonumber \\
 = \int_{\partial U} d^{n-1}u \sqrt{h}\; \frac{P(\bm{u}|J)}{\sqrt{g}}n_\sigma v^\sigma = 0.
\end{eqnarray}
Defining the expectation $\big<A\big>_J$ as 

\begin{equation}
\Big<A(\bm{u})\Big>_J = \int_U d^nu P(\bm{u}|J)A(\bm{u}),
\end{equation}
we arrive at the identity

\begin{equation}
\Big<\partial_\mu v^\mu\Big>_J + \Big<v^\mu \partial_\mu \ln P(\bm{u}|J)\Big>_J = 0.
\end{equation}

This can be written in a manifestly covariant manner as

\begin{equation}
\Big<\nabla_\mu v^\mu\Big>_J + \Big<v^\mu\partial_\mu \ln \frac{P(\bm{u}|J)}{\sqrt{g}}\Big>_J = 0.
\end{equation}

\section*{Acknowledgment}

This work was funded by FONDECYT grant number 1140514 and partially funded by CONICYT ACE-01 and PIA-CONICYT ACT-1115 grants.
The author gratefully acknowledges conversations with L. Vel\'azquez at UCN, Antofagasta regarding the
role of differential geometry in probability theory, and Ariel Caticha on the foundations of entropic inference.

\bibliographystyle{ieeetr}
\bibliography{maxent}

\end{document}